# SEGNET: Secure Geo-Sensor Network Model

Tapalina Bhattasali

*Department of Computer Science & Engineering, University of Calcutta, Kolkata, India*
tapolinab@gmail.com

*Abstract*— Wireless Geo-Sensor Networks (GEONET) are suitable for critical applications in hostile environments due to its flexibility in deployment. But low power geo-sensor nodes are easily compromised by security threats like battery exhaustion attack which may give rise to unavoidable circumstances. In this type of attack, intruder forcefully resists legitimate sensor nodes from going into low-power sleep state. So that compromised sensor nodes' battery power are drained out and they stop working. Due to the limited capability of sensor nodes, it is very difficult to prevent a sensor node from this type of attack which apparently appears as innocent interaction. In this paper, a framework of secure GEONET model (SEGNET) is proposed, based on dynamic load distribution mechanism for heterogeneous environment. It considers hybrid detection approach using three modules for anomaly detection, intrusion confirmation and decision making to reduce the probability of false detection, compared to other existing approaches.

## I. INTRODUCTION

Nowadays, natural hazards are increasing due to various reasons such as global warming, climate change etc. The losses due to these hazards are increasing at an alarming rate. However, these environmental hazards are largely unpredictable and occur within very short spans of time. Therefore, technology has to be developed to capture relevant signals with minimum monitoring delay. Wireless geo-sensor network (GEONET) is one of the cutting edge technologies that can quickly respond to the rapid changes of sensed data in the surrounding environment.

Most of the deployed wireless geo-sensor networks are used to measure scalar physical phenomena such as temperature, humidity etc. The low cost and small size of the sensors, coupled with their ability to communicate without any infrastructural support makes them a necessity in crisis management. GEONET is mainly used for environment monitoring such as coastal monitoring, ocean exploration, flood management, forest fire detection, habitat monitoring etc. GEONET is useful in environmental hazard management by providing a system which will be able to learn about the phenomena of natural hazards and will provide early warning signal [1]. In this field, accuracy and response time play significant role. Therefore, the security of sensor networks becomes very important. But applications of GEONET become useless, if the delay due to anomaly is too large. Incorrect or unavailable query results may cause serious damage.

GEONETs are much more vulnerable to battery exhaustion attack than the conventional networks due to the limited capability of sensor nodes and the lack of centralized monitoring and management in sensor networks. Target of battery exhaustion attack [2] is to maximize the power consumption of the affected node, thereby decreasing its battery life. Sleep deprivation attack is the most devastating one. Maximum security can only be achieved by designing secure GEONET model whose purpose is to provide alert about possible intrusion, ideally in time to stop the attack or to mitigate the damage. Therefore focus of this paper is to propose a secure lightweight framework which can detect intrusion accurately in time so that no data becomes inaccurate or delayed.

The remainder of this paper is organized as follows. Section II explains the problem domain. Section III describes related works and section IV consists of an outline of the proposed system model. It is followed by a conclusion in section V.

## II. PROBLEM DOMAIN

According to Marco Conti et. al.[3] wireless sensor networks represent a special class of multi-hop ad hoc networks that are developed to control and monitor events and phenomena where research works are still expected to address specialized problems ranging from QoS to privacy, security and trust, specialized network scenarios, or the usage of sensor networks in challenging environments.

The nature of GEONET deployment in severe geographical terrains makes it impossible to recharge the battery power of sensor nodes. Quick battery depletion leads to death of sensor nodes and eventual shut-off occurs in the network either fully or partially. The absence of infrastructure also makes it difficult to detect security threats, often resulting in reduced Quality of Service (QoS). If a monitoring sensor node gets affected by intruder, then the node can behave abnormally and alarms might be generated at wrong times. This may lead to incorrect decisions in environmental hazard management. When a relay node is affected by attack, transmission of important information may get blocked or may not take place. As a consequence, rescue operations get delayed and emergency situations may develop. Therefore, security solutions in GEONET are vital for the wider acceptance and use of sensor networks and have to be designed with efficient resource utilization. The need of the day is to design a model for detecting intrusions accurately in an energy-efficient manner. For this reason, a system model SEGNET is proposed to extend the lifetime of GEONET, even in the face of battery exhaustion attack.

## III. RELATED WORKS

Intrusion detection can be implemented using various techniques. Intrusion detection for GEONET is an emerging field of research. In one of the previous papers [4], a survey of recent IDS in sensor network has been presented. This section presents a review of on-going research activities in this context.

In one of the earlier works, a lightweight hierarchical model [5] is proposed for heterogeneous wireless sensor network to detect insomnia of sensor nodes. In this model, six types of component nodes are available at five layers. Sector coordinator collects data from leaf nodes and anomalous packets are transmitted to sector monitor and valid packets are transmitted to forwarding sector head. Valid packets are forwarded to cluster coordinator and if it is really valid then forwarded to sink node, otherwise dropped. If anomalous packets are really suspected then it is dropped by sector monitor, otherwise forwarded to cluster coordinator. It uses two phase anomaly detection technique to avoid phantom detection. It focuses on intrusion detection at layer 1.This type of model can extend the network lifetime even in the face of sleep deprivation attack, but it enhances packet overhead.

In another approach, a mathematical model [6] based on Absorbing Markov Chain (AMC) is proposed for Denial of Sleep attack detection in sensor network. In this mechanism, whether sensor network is affected by denial of sleep attack or not can be decided by considering expected death time of sensor network under normal scenario. As behavior of sensor nodes varies with time, proposed intrusion detection model in wireless sensor network is probabilistic one. It works more accurately than deterministic model. But it considers only the overall network not the individual node.

Therefore to improve the performance of intrusion detection methodology, SEGNET model is proposed in the next section. This model is a modified version of one of the earlier works [7] for detecting sleep deprivation attack.

## IV. PROPOSED MODEL

In this section, a novel framework of Secure GEONET (SEGNET) model has been proposed for detecting intrusion in wireless geo-sensor network. It uses cluster based mechanism [8] in an energy efficient manner that enhances network scalability and lifetime. In this model, energy efficiency can be achieved by using load distribution mechanism where low energy nodes are assigned the task of sensing. Dynamic model [9] is designed here to overcome sudden death of sensor nodes. In SEGNET model, compromised sensor nodes are accurately identified and isolated, if it is required.

In the proposed model, nodes at different layers are categorized into three different types depending on their battery capacity; (i) Base node (ii) Intelligent node (base nodes which are not attached to micro-server),(iii) Simple node (having only sensing capacity). Depending on the functionality, sensor nodes are categorized into the following designations such as Gateway Node (GN), Cluster Owner (CO), Monitor Node (MN), Zone Owner (ZO) and Sensing Node (SN). There are several types of sensing nodes for sensing temperature, humidity, light, pressure, rainfall, wind speed etc.

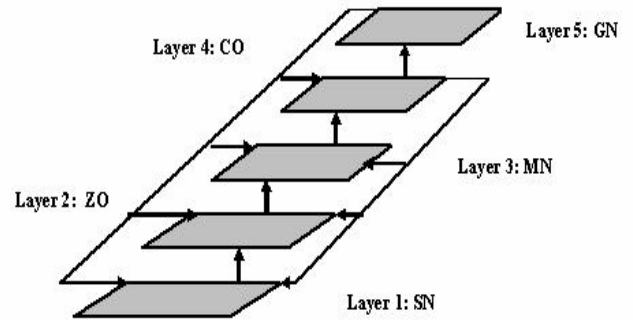

Fig. 1 Hierarchical SEGNET Model

Fig 1 represents the hierarchical model. Participant nodes in SEGNET model are defined here depending on their functionality.

- GN: One type of base node having high capacity among other nodes.

- CO: One type of intelligent node which has maximum energy level, degree (number of nodes within its coverage area) and minimum distance among all neighbors of GN. It acts as the controller of cluster area. It is capable to take final decision regarding intrusion.

- MN: One type of intelligent node who has minimum distance from cluster owner and maximum energy among all neighbor nodes of cluster owner. It is responsible for checking data flow to analyze network traffic and to confirm intrusion.

- ZO: One type of intelligent node whose degree is highest among all the neighbors of CO. It is capable to collect sensing data from SN and to detect anomalous behavior if any.

- SN: One type of simple node for sensing events. Its detection power is disabled.

Some of the assumptions of SEGNET model are given below.
- Network is divided into clusters which are again partitioned into zones.
- Gateway Node acts as honest gateway to another network or access point.

- Energy contents $E_{IN}$ of intelligent nodes have µ times more energy than simple nodes.

The life-cycle of SEGNET model is divided into the following phases.

*1) Initialization Phase*

Geo-Sensor nodes are deployed. GN broadcasts HELLO MESSAGE and its node-id and designation. GN sends query message to acquire energy status from any node N joining the network and categorizes it according to its response. GN sets the node N's id and other parameters.

COs are selected. CO adds nodes within its coverage area into its member list. Then MNs are selected for each cluster. ZOs are selected and add nodes within its coverage area into its member list. Pre-loaded detection modules are activated according to the node's designation.

Candidates can be selected randomly if more than one fulfills the selection criterion.

*2) Data Collection Phase*

When query request comes from GN [10], corresponding sensing node in sleep state receives wake-up coin [11] from its ZO who collects sensing data packets and checks for anomaly. If anomaly is detected, corresponding packet is marked as suspected.

*3) Data Transfer Phase*

ZO sends packets towards CO. MN checks for intrusion during traffic flow from ZO to CO. If real intrusion is detected, warning ticket is sent to CO for each data packet. If warning tickets are generated for a packet from at least two different MNs, packet is considered as fake and rejected. If number of warning tickets received within a time interval is greater than threshold, sensing node is blocked. If no intrusion is detected, packet is forwarded to GN.

If GN discovers any new node N in the middle of duty cycle, it sends sleep signal to node N.

When any sensor node's behavior deviates from normal behavior, reconfiguration procedure takes place or, after a pre-defined time-interval, reconfiguration procedure takes place to avoid complete exhaustion of sensor nodes. During reconfiguration, detection modules of the previously selected nodes are disabled and detection modules of currently selected nodes are enabled.

When network lifetime reaches below threshold value, GEONET is deactivated.

*A. Case Study of SEGNET Model Functionality*

Fig. 2 represents a case study of SEGNET model. It consists of number of sensor nodes such as A,B,C,D,E,F,G,H,I,J,K,L,M,N. Here node A, node B, node C and node D are designated as Sensing Node (SN). Node E and node F are designated as Zone Owner (ZO). Node G, node H, node I, node J, node K and node L are designated as Monitor Node(MN). Node M is designated as Cluster Owner (CO) and node N is designated as Gateway Node (GN).

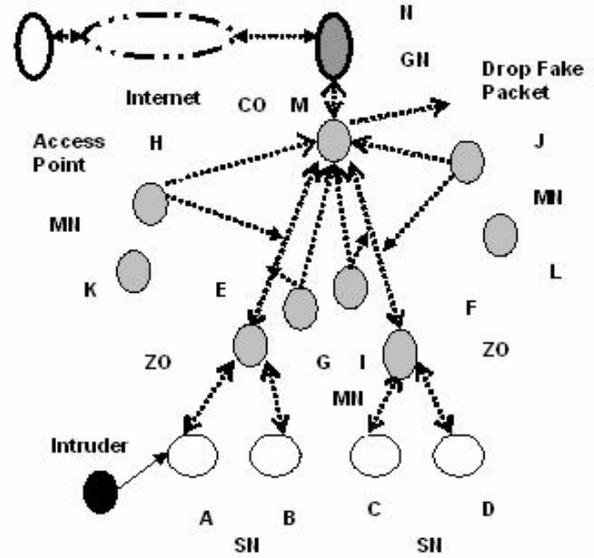

Fig. 2   Working Principle of SEGNET Model

Assume the case where one unknown node (intruder) sends fake requests to Sensing Node (SN) A i.e. node A is compromised. Suppose, Gateway Node (GN) sends query through Cluster Owner (CO) node M towards Zone Owner (ZO) node E. Then node E e.g., sends wake-up coin to node A. As a result node A sends packet P1 to ZO node E. If any anomaly is detected by node E, it sets status of packet P1 to 1(suspected). Node E forwards packet P1 to CO node M. During transmission, monitor nodes (MN) G and H checks traffic for intrusion. If intrusion is detected, warning ticket is sent to CO node M. If node M receives warning tickets from both G and H, then packet P1 is considered as fake packet which is ignored and node A is kept under observation. If warning tickets generated for node A from different MNs within a time interval exceeds threshold value, then node A is blocked.

Suppose data need to be transmitted from node D at next time. ZO node F collects data, checks for anomaly. If no anomaly is found, status of packet p2 is set to 0. Data is transmitted towards CO node M when traffic is analyzed by monitor nodes I and J. If no intrusion is found, no warning ticket is generated , CO sends data towards GN node N.

*B. Procedural Logic of SEGNET*

TABLE I
DATA DICTIONARY

| Parameters | Description |
|---|---|
| G_neibor { } | Set of neighbor nodes of GN. |
| $E_N$ | Energy level of node N. |
| $E_{GN}$ | Energy level of GN node. |
| Intelligent_Nd{ } | Set of intelligent nodes. |
| Simple_Nd{} | Set of simple nodes. |
| desig($N_i$) | Designation of node $N_i$. |
| Distance$_{N,CO}$ | Distance between node N and CO. |
| MIN (Dist_Cneibor{}) | Minimum distance from CO among all neighbors of it. |
| MAX (Eng_Cneibor{}) | Maximum energy level among all neighbors of CO. |
| degree(N) | Number of nodes within the range of node N. |
| MAX (Deg_Cneibor{}) | Maximum degree among all neighbors of CO. |
| ZO_neibor{} | Set of neighbors of ZO. |
| MAX (Deg_Gneibor{}) | Maximum degree among all neighbors of GN. |
| Res_Eng(N) | Residual energy of node N. |
| MAX (RE_Gneibor{}) | Maximum residual energy among all GN neighbors. |
| C_neibor{} | Set of neighbors of CO. |
| SneiborCO{} | Set of simple node neighbors of CO. |
| cnt(wakup(tinter)) | Number of wakeup coin received within time interval. |
| Th_Token | Threshold number of tokens. |
| Th_max | Maximum threshold number of packets. |
| Th_min | Minimum threshold number of packets. |
| Th$_{ENERGY}$ | Threshold energy level. |
| cnt(Warning) | Number of warning tickets generated by one MN within specific time interval. |
| maturity | Experience level 0 implies no experience and 1 implies experienced. |

*Main Module*
1. GN broadcasts its profile and starts timer T.
2. If acknowledgement from N is received within timeout then, add node N to G_ neibor{}.
3. For every N∈G_ neibor{}, GN collects energy $E_N$.
4. If $E_N \neq (1/\mu)* E_{GN}$ then,
    $N_i$ is added to Intelligent_Nd{ }.
   Else $N_i$ is added to Simple_Nd{}.
5. CO_Select ().
6. Cluster_Form().
7. // Selection of Monitor Node
   If Distance$_{N,CO}$ = MIN(Dist_Cneibor{})
       and $E_N$ = MAX(Eng_Cneibor{}) then
       Set desig($N_i$) as MN.
8. // Selection of Zone Owner
   If degree($N_i$) = MAX(Deg_Cneibor{})
       Set desig($N_i$) as ZO.
9. // Zone form within coverage area of ZO
       ZO collects its neighbor ids into ZO_neibor{}.
10. Collect_Info().
11. Forward Packets to CO.
12. Confirm_Intrusion ().
13. Action().
14. ENDFOR
15. End

*1) CO_Select()*
1. Check whether neighbor node $N_i$ is type of intelligent node or not.
   If $N_i \in$ Intelligent_ Nd{ } then,
       $N_i$ is added to candidate_CO{ }.
   // candidate_CO{ } consists of all candidates of CO.
2. Compute degree of each neighbor node of GN.
3. If degree ($N_i$)>= MAX(Deg_GNeibor{}) then,
   If Res_Eng(N)>=MAX (RE_GNeibor{}) then,
       prob($N_i$) =1.
       //$N_i$ has the probability of becoming CO.
   Else exit.
4. If maturity of the node $N_i$ equals to 0 then,
       desig($N_i$) is set to CO.
5. Set maturity of the node is to 1.
6. Broadcast profile of new CO.

*2) Cluster_Form()*
   // Procedure to form cluster within coverage area of CO.
1. CO broadcasts its profile and starts timer T.
2. Nodes that acknowledge within this period are added to C_neibor{}.
3. Simple nodes are added to SneiborCO{}.

*3) Collect_Info()*
   // Step to collect data from sensing leaf node.
1. GN sends query for sensing data to ZO through CO.
2. ZO sends wake-up coin to corresponding $N_i$ and computes current energy level of Node .
3. ZO calls Anomaly_Detec().

*4) Anomaly_Detec()*
   // Step to detect anomaly during data transmission.
1. /* check whether packet received from $N_i$ during sleep schedule of $N_i$ */
   If $T_{SLP(START)} <= T1 <= T_{SLP(END)}$ then,
       /* where $T_{SLP(START)}$ and $T_{SLP(END)}$ represent begin and end of sleep schedule for node $N_i$ and receipt time(packet($N_i$))=T1*/
2. /*count number of wake-up coin received by $N_i$ within a specific time interval*/
   If count(wakeup(tinterval))> Th_Token then ,
       anomalous event occurs.
3. ZO sets status(pkt$_i$)=1(suspected) to any packet from $N_i$ , otherwise status(pkt$_i$)=0 (genuine).

*5) Confirm_Intrusion()*
// During transmission, MN monitors traffic whether intrusion occurs or not.
1. If Th_max <count($p_i$ from $N_i$ )< Th_min then,

check, if Res_Energy($N_i$) <$Th_{ENERGY}$ then,
Warning ticket generated by MN to CO.

*6) Action()*

1. If any packet is received by CO, whose status is set to 1, but no warning is generated then,
   packet is erroneous but no intrusion.
2. If CO receives warning tickets (WT) from more than one MN for the same packet then,
   CO ignores corresponding fake packet irrespective of its status field.
3. Related valid data packets are aggregated and forwarded to GN.
4. If count (Warning)> Threshold then,
   $N_i$ is blocked for further communication.
   Else
   $N_i$ remains under observation.

*C. Additional Features of SEGNET Model*

Although SEGNET model focuses on intrusion detection at low capacity sensing nodes, detection can also be possible at other nodes in GEONET.

- If ZO repeatedly sets valid packets as suspected i.e, when count(false detection) > threshold, it is considered as compromised and MNs send warning tickets to CO. Then old one is blocked and new ZO is selected.
- If warning ticket rate (the ratio of the number of reported warning tickets from distinct MNs to the total number of MNs in the cluster) exceed predefined threshold within a time interval then compromised MN is blocked and new one is selected.
- If MNs (more than one) detect abnormal flow volume of CO, they send warning tickets to GN. Then compromised CO is blocked and new one is selected.

To prove the efficiency of the proposed model SEGNET, performance of the model needs to be evaluated through simulation. Work is going on to analyze performance of the proposed SEGNET model.

## V. CONCLUSION

Wireless geo-sensor networks are susceptible to a variety of security threats due to its deployment in open and unprotected domain. As battery exhaustion attack has the power to quickly cut-off parts of the geo-sensor network by exhausting the energy levels of compromised nodes, early detection is important. In this paper, an effort has been made to propose a collaborative model capable of detecting intrusion. The aim of proposed SEGNET model is to save the energy of sensor nodes so as to extend the lifetime of the network, even in the face of attack. Proposed model virtually eliminates the probability of phantom detection [12] by using three step processes. Workload of SEGNET model is distributed among the components according to their capacity to avoid complete exhaustion of battery power. More studies are being undertaken to analyze the performance of the proposed model and will be compared with other existing models.